\documentclass{aa}  

\usepackage{graphicx}
\usepackage{txfonts}
\usepackage{subcaption}
\usepackage{hyperref}
\hypersetup{colorlinks=true,citecolor=blue}
\newcommand{\flowrate}{~M$_{\sun}\mathrm{yr}^{-1}$}
\begin{document}

   \title{From theory to observation: understanding filamentary flows in high-mass star-forming clusters}
   \author{M. R. A. Wells\inst{1,2} \and R. Pillsworth\inst{3} \and H. Beuther\inst{1} \and R. E. Pudritz\inst{3,4} \and E. W. Koch\inst{5,6}}
   
   \institute{Max Planck Institute for Astronomy, Königstuhl 17, 69117 Heidelberg, Germany \and Fakultät für Physik und Astronomie, Universität Heidelberg, Im Neuenheimer Feld 226, 69120 Heidelberg, Germany \and Department of Physics and Astronomy, McMaster University, Hamilton, ON L8S 4M1, Canada \and Origins Institute, McMaster University, Hamilton, ON L8S 4M1, Canada \and Center for Astrophysics, Harvard \& Smithsonian, 60 Garden Street, Cambridge, MA 02138-1516, USA \and National Radio Astronomy Observatory, 800 Bradbury SE, Suite 235, Albuquerque, NM 87106, USA}

   \date{}

  \abstract{Filamentary structures on parsec scales play a critical role in feeding star-forming regions, often acting as the main channels through which gas flows into dense clumps that foster star formation. Understanding the dynamics of these filaments is crucial for explaining the mechanisms of star formation across a range of environments.}{Here we use data from multi-scale galactic MHD simulations to observe filaments and star forming clumps on 10's of pc scales and investigate flow rate relationships along, and onto filaments as well as flows towards the clumps.}{Using the \texttt{FilFinderPPV} identification technique, we identify the prominent filamentary structures in each data cube. Each filament and its corresponding clump are analysed by calculating flow rates along each filament towards the clump, onto each filament from increasing distances, and radially around each clump. This analysis is conducted for two cubes, one feedback dominated region, and one with less feedback, as well as for five different inclinations (0, 20, 45, 70, and 90 degrees) of one filament and clump system.}{Looking at the face-on inclination of the simulations (0 degrees), we observe different trends depending on the environmental conditions (more or less feedback). The median flow rate in the region with more feedback is 8.9$\times$10$^{-5}$\flowrate and we see that flow rates along the filaments toward the clumps generally decrease in these regions. In the region with less feedback we have a median flow rate of 2.9$\times$10$^{-4}$\flowrate and when looking along the filaments here we see the values either increase or remain constant. We find that the flow rates from the environments onto the primary filaments are of an order of magnitude sufficient to sustain the flow rates along these filaments. When discussing the effects of galactic and filamentary inclination, we also observe that viewing the filaments from different galactic inclinations can reveal the presence of feeder structures (smaller filamentary structures aiding in the flow of material). Additionally, considering the inclination of the filaments themselves allows us to determine how much we are overestimating or underestimating the flow rates for those filaments.}{The different trends in the relationship between flow rate and distance along the filaments in both the feedback and non-feedback dominated cubes confirm that the environment is a significant factor in accretion flows and their relationship with filament parameters. The method used to estimate these flow rates, which has been previously applied to observational data, produced results consistent with those obtained from the simulations themselves, providing high confidence in the flow rate calculation method.} 

   \maketitle
\section{Introduction}

\begin{figure*}
    \centering
    \includegraphics[width=1.0\linewidth]{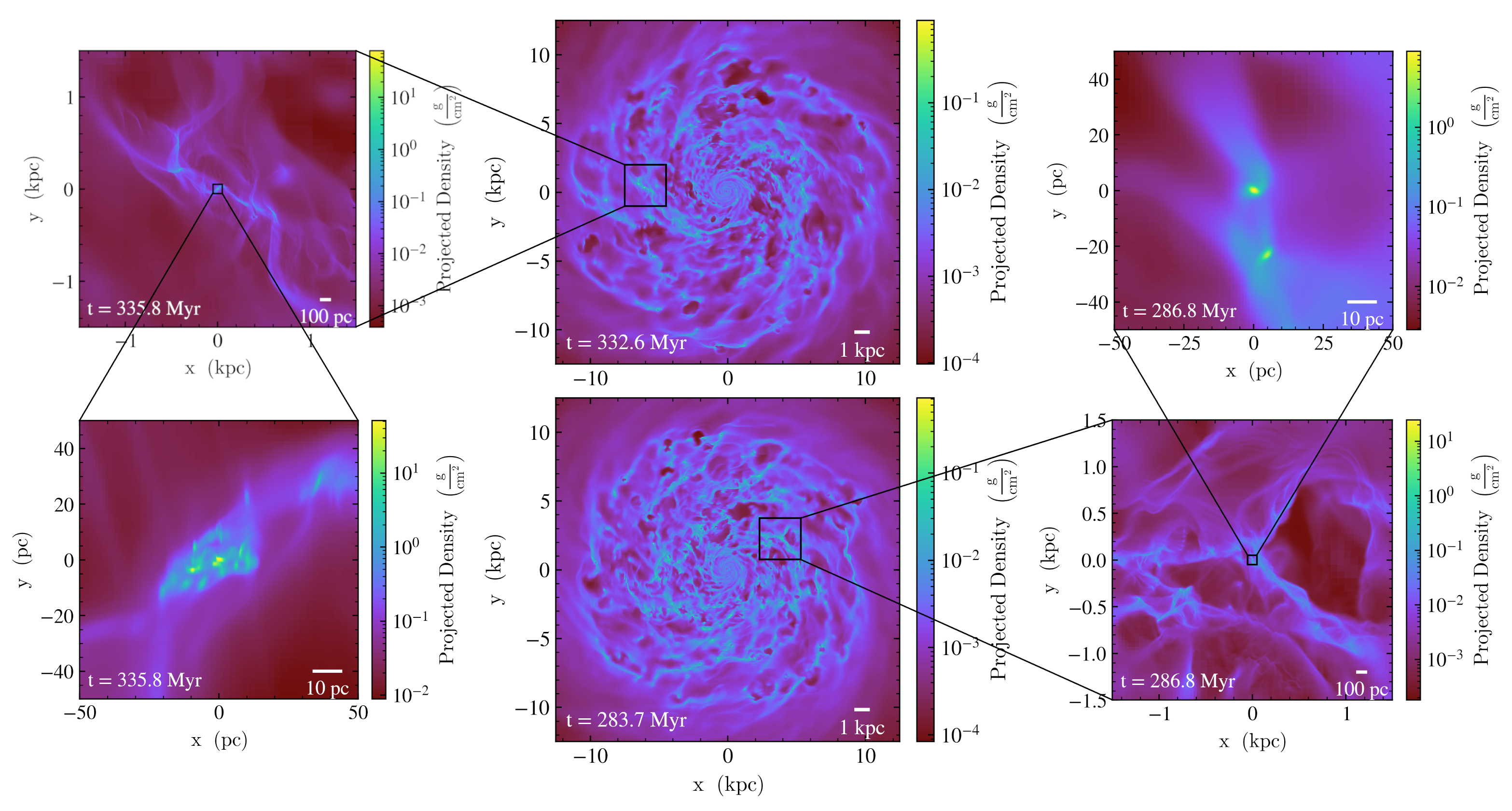}
    \caption[Simulated galaxy overview]{Galaxy overview from \citet{ZhaoPudritz2024}. The central panels show the two snapshots of the galaxy we are using, the top showing the location of the less feedback dominated region (quiet) and the bottom showing the location of the feedback-dominated region (active). The first zoom-in panels show the regions down to a few kpc (top left and bottom right panels), followed by the close ups of the regions in 100 x 100 pc boxes (bottom left and top right panels).}
    \label{fig:largescale}
\end{figure*}

Giant Molecular Clouds (GMCs) serve as essential structures within galaxies, acting as intermediaries (on the order of several tens of parsecs) that connect large-scale galactic dynamics to the smaller-scale localised star formation processes. 
The collapse and fragmentation of GMCs into dense regions capable of star formation is driven by a combination of gravitational instabilities and external pressures from the surrounding interstellar medium (e.g.,~\citealt{1984MZinnecker, 2003Bonnell, 2014Andre, 2018Urquhart, 2019Svoboda, Padoan2020} ). As these clouds cool and accumulate mass, they fragment into smaller, denser regions under gravitational contraction. With extreme temperatures and pressures these smaller dense regions continue to collapse, forming clusters of protostellar objects. 
Intersections within these filamentary networks, known here as “hubs”, often serve as sites for the formation of high-mass stellar clusters, where the convergence of gas flows provide perfect conditions for the majority of stellar births (e.g,~\citealt{Lada_2003, 2008Goldsmith, 2009Myers, 2010Andre, 2010Schneider, 2010Bressert, 2013Kirk, 2014Krumholz,  2020Kumar, 2021Grudic, 2025Hacar}). We see a self-similarity among the scales here too, where GMC's are hub sites on larger scales too (e.g,~\citealt{2024ZDD}).
These filaments and filamentary-like structures have been a part of the discussion for years in different shapes and forms, looked at in different tracers, and as a part of many different studies both theoretically and observationally (e.g.,~\citealt{2000Fiege, 2010Jackson, 2013Kirk, 2014Gomez, 2014Henshaw, 2018Chira, Padoan2020, 2020Alves, 2020Beuther, 2020Schisano, 2023Hacar, 2024Pillsworth, 2024Wells}).
Evidence for the feeding of clouds, clumps and cores being done by filamentary structures can be seen in many of the studies mentioned above. These studies highlight how filaments, ranging in scale from galactic kpc scales down to sub-parsec levels, connect the parental molecular clouds, cluster forming hubs, clumps and individual cores, demonstrating the critical role they play in channelling mass and angular momentum. 
Despite the progress made in understanding filamentary structures and their role in star formation, several questions remain. The precise mechanisms by which material is transported through, along and around filamentary networks are still not well defined and their impact on the formation of high-mass star clusters are still not completely understood. Current simulations and observations continue to challenge our understanding of these processes, emphasising the need for further research to unravel the complexities of filament dynamics and their contributions to stellar cluster formation.

The advancement of theoretical model capabilities combined with large mm/sub-mm interferometers, such as the Atacama Large Millimeter/submillimeter Array (ALMA), the Northern Extended Millimeter Array (NOEMA) and the Submillimeter Array (SMA), has allowed in-depth research into more complex galactic structures such as filaments, on multiple scales. Observational (e.g.,~\citealt{2014Ragan, 2015Zucker, Russell_2017, 2018Hacar, 2019Olivares}) and computational (e.g.,~\citealt{2014Gomez, 2016Federrath, 2019Haid, 2019Li, Padoan2020, ZhaoPudritz2024}) studies are complementing each other, using observational constraints in models, or theoretical limitations in observational analysis which allow the field to progress further (e.g.,~\citealt{2014Clark, Hillel_2020, Duan2024}). This collaborative approach is often underscored in reviews of the field, such as those by \citealt{2014Andre, 2023Pineda}, which emphasise the importance of combining theoretical and observational perspectives. Together, these methodologies are advancing our understanding of the nature of filaments, allowing researchers to investigate their dynamics and trace their evolution in unprecedented detail.

In this paper, we use data cubes from the simulations by \cite{ZhaoPudritz2024}. For our observational approach, we use position-position-velocity (PPV) data drawn from these simulations - chosen at 5 different inclination angles (0, 20, 45, 70 and 90 degrees, assuming face on is 0) and for 2 different environments (more and less feedback). We measure flow rate properties along filaments, onto filaments and radially onto cluster forming clumps. Here we are investigating larger scale relations between flow rates and filamentary parameters, environment, and inclination. A major feature of this work is that we also test the validity of observational methods by comparing the calculated values from our observational method with those from the full numerical PPP data from the simulation - information that is not available from observations of real systems.

The structure of the paper is as follows: the simulation data is introduced in Sect. \,\ref{sect:theorydata}. In Sect. \,\ref{sect:filfind}, we introduce \texttt{FilFinder}, the package used to identify filaments in the PPV data cubes, and discuss the different parameters. Details of how perpendicular, parallel and polar flow rates in these regions are calculated are presented in Sect. \,\ref{sect:theorymethod}, before showing the results of the observational method in Sect. \,\ref{sect:theoryresults}. The results of the analysis of the flow rates deduced from the 3D simulation PPP data, and their comparison to the observational method, are then presented in Sect. \ref{sect:results_simulations}. Finally in Sect. \,\ref{sect:theorydiscussion}, we discuss the results including other scales and comparisons with observations from several programs before we draw our conclusions and discuss opportunities for future work in Sect.\,\ref{sect:theoryconc}.

\begin{figure*}[ht]
\begin{subfigure}{0.5\textwidth}
\includegraphics[width=0.99\textwidth, height=0.28\textheight]{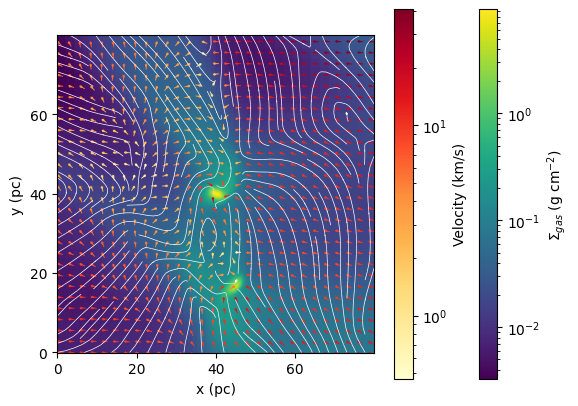} 
\caption{Active region, taken at a snapshot of $\simeq$ 283 Myr.}
\label{fig:activecube}
\end{subfigure}
\begin{subfigure}{0.5\textwidth}
\includegraphics[width=0.99\textwidth, height=0.28\textheight]{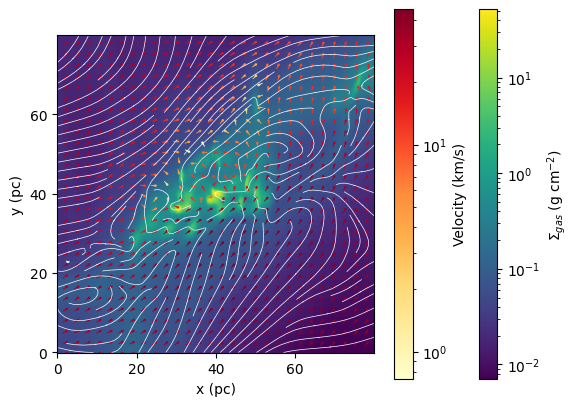}
\caption{Quiet region, taken at a snapshot of $\simeq$ 332 Myr.}
\label{fig:quietcube}
\end{subfigure}
\caption[Active and quiet region density projections]{Density projections of high-resolution zoom-in simulation data of \citet{ZhaoPudritz2024}. White streamlines represent magnetic field structure. Quivers show velocity direction and magnitude relative to the velocity of the central cores in each snapshot.}
\label{fig:theorycubes}
\end{figure*}
\section{Simulation Data}
\label{sect:theorydata}

We use data from multi-scale MHD simulations of a Milky Way type galaxy from \citet{ZhaoPudritz2024}. Those simulations were run in \textsc{ramses} with the AGORA project initial conditions \citep{KimAgertz2016a}. These include a dark matter halo with M\textsubscript{DM halo} = 1.074 x $10^{12}$ M\textsubscript{\(\odot\)}, R\textsubscript{DM halo} = 205.5 kpc, and a circular velocity of v\textsubscript{c,DM halo} = 150 kms$^{-1}$, an exponential disk with M\textsubscript{disk} = 4.297 x $10^{10}$ M\textsubscript{\(\odot\)}, and a stellar bulge with M\textsubscript{bulge} = 4.297 x $10^{10}$ M\textsubscript{\(\odot\)} (\citep{KimAgertz2016a}). For further simulation details we refer to \citet{ZhaoPudritz2024, KimAgertz2016a}. 

Recently, \citet{Pillsworth2025} characterised the properties of over 500 galactic scale filaments in the \citet{ZhaoPudritz2024} simulations using the Filfinder package \citep{2015Koch}. That work derived the mass distribution function and gravitational stability of filaments but did not investigate their flow dynamics.

This paper focuses specifically on the so-called active and quiet zoom-in regions within the galactic disk of \citet{ZhaoPudritz2024}.
Figure \ref{fig:largescale} shows this galaxy in the central panels, with the two regions used in this work marked on top, and zoom panels shown either side. The active region (right panels) is dominated by feedback, in an area of converging super bubbles, whereas the quiet region (left panels) has less feedback and is in a spiral arm like area of the galaxy. These zoom-in regions are 3~kpc wide boxes around dense proto-clusters that achieve a spatial resolution of up to 0.28 pc.  We extract data from 60~pc around the densest cell to focus on the star-forming cluster.  

In Fig. \ref{fig:theorycubes} we show the two regions in projected density. Arrow quivers indicate the rotation-corrected velocity flow in the plane on projection, while streamlines track magnetic field structure. These figures outline the larger-scale gas flows. In panel (a), the feedback-dominated region, one sees that the gas flows onto the filamentary structures. This is caused by the feedback of the surrounding super bubbles. The velocities there have ordered gradients with clear increasing trends, direction pointing towards the central clump. In panel (b), the non-feedback region, on larger scales the velocities are almost parallel to the dense filament whereas on small scales the velocity field appears more chaotic leading to the many clumps. Looking at the magnetic field lines displayed in white we see they are more disordered in the region with less feedback (panel (b)), where as in panel (a), where we have more feedback, the lines follow parallel to the filament structure.

The extracted cubes are run through a position-position-velocity (PPV) post-processing code, using YT \citep{TurkSmith2011} and the YT astro analysis extension \footnote{Astro analysis code here: \url{https://github.com/yt-project/yt_astro_analysis}}. This step is key for the observational comparison as observations only produce PPV data cubes and not position-position-position (PPP). The cubes are 212 x 212 pixels at a resolution of 1 px = 0.285 pc with velocity channels at resolution of 0.8~kms$^{-1}$. In terms of how this compares to resolutions we see in observation data, the velocity resolution is of a similar value to what was used in \cite{2024Wells}. Spatially we have a likeness to larger scale single-dish data. Since we leave the galactic rotation to be corrected at a later point in the analysis, the bounds of the velocity channels change with inclination as the rotation of the galaxy becomes more dominant. For the face on (0 degrees) cubes we have a velocity range of -19.6~kms$^{-1}$ to 19.6~kms$^{-1}$. The PPV processing does not include any spectral line post-processing on the data, and instead returns column densities of the areas in cm$^{-2}$.


\section{Methods}
\label{sect:theorymethod}
\begin{figure*}[ht]
\begin{subfigure}{0.5\textwidth}
\includegraphics[width=0.9\textwidth, height=0.3\textheight]{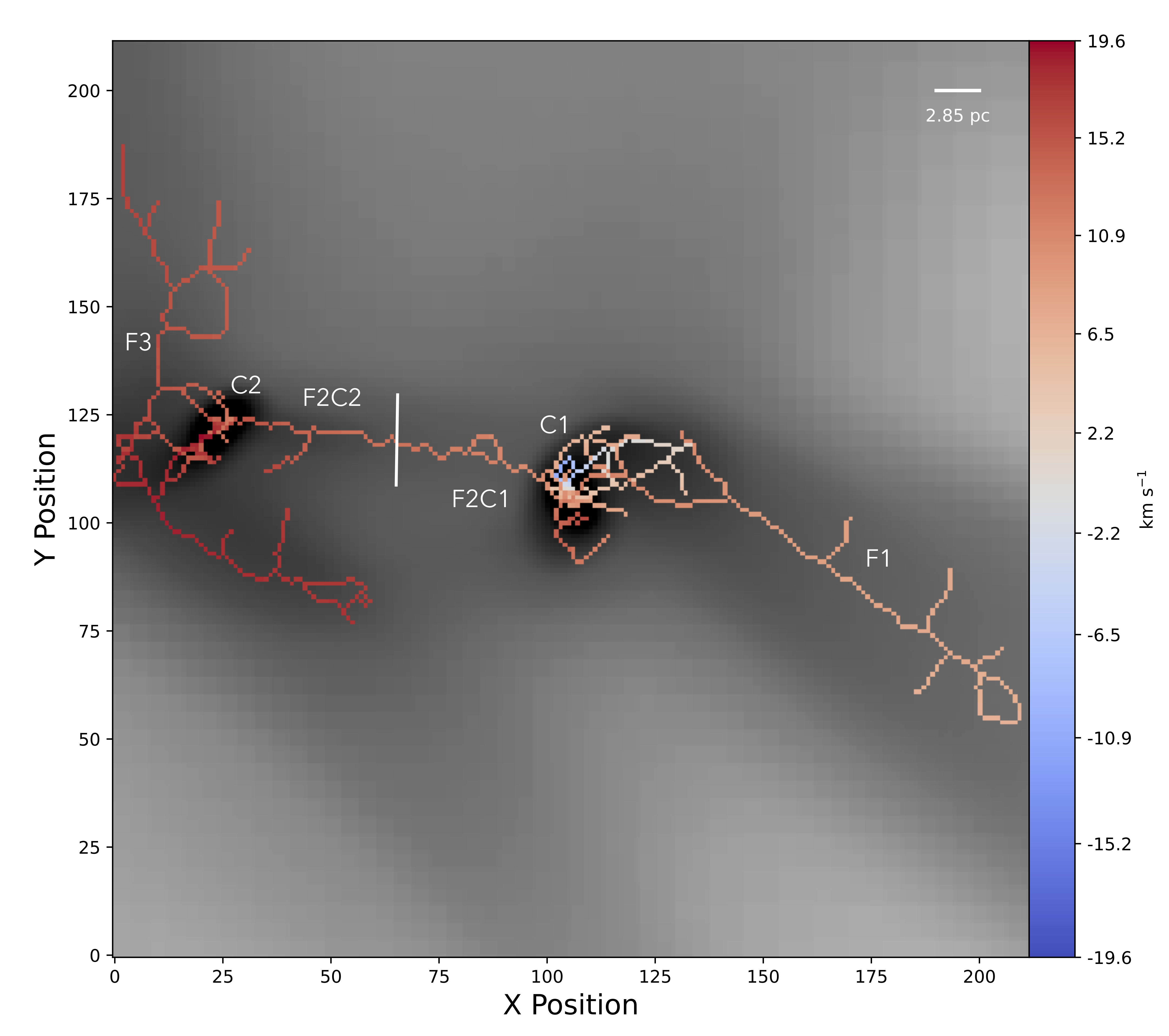} 
\caption{Active (feedback-dominated) region}
\label{fig:activecubeskel}
\end{subfigure}
\begin{subfigure}{0.5\textwidth}
\includegraphics[width=0.92\textwidth, height=0.3\textheight]{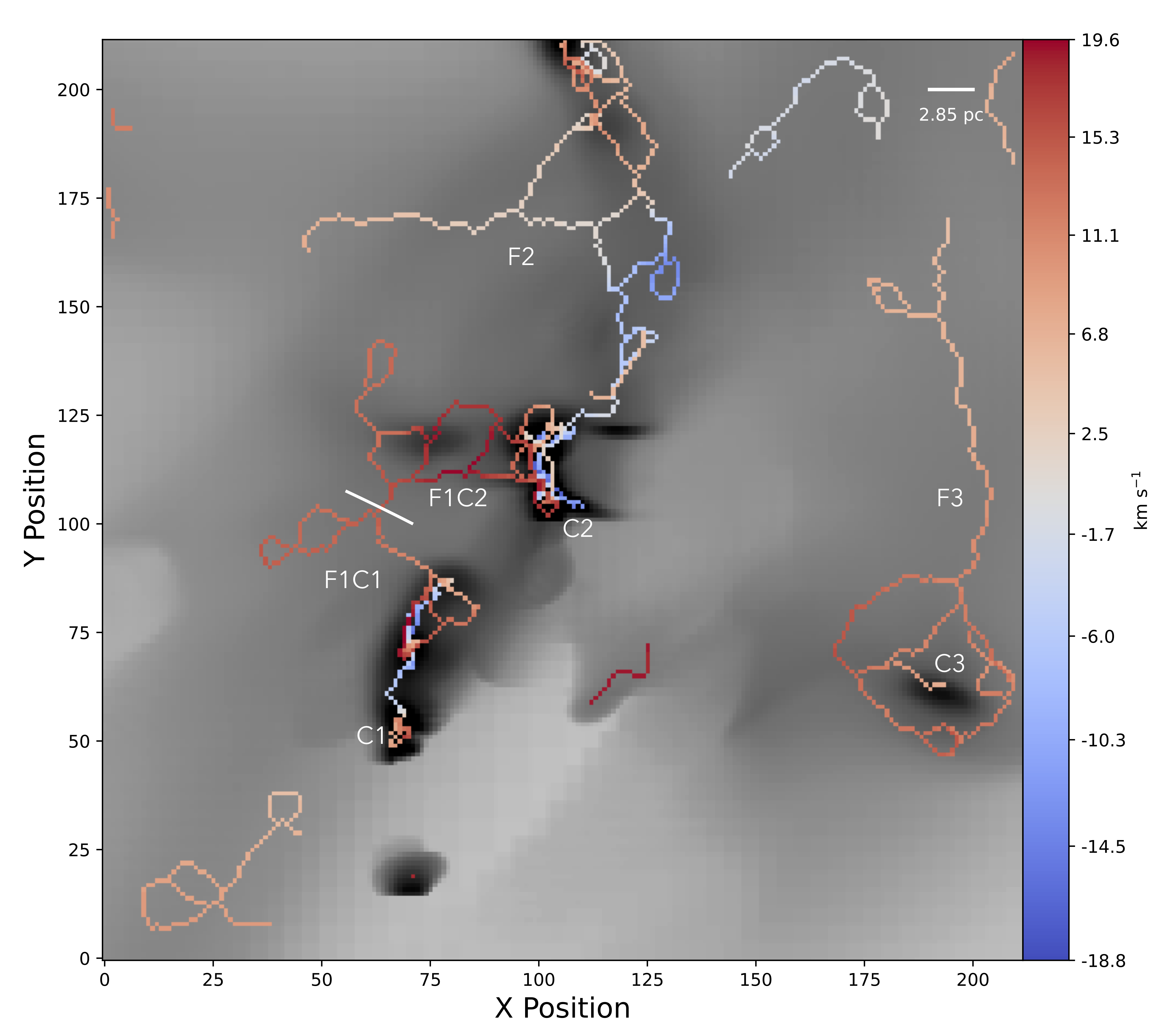}
\caption{Quiet (non feedback-dominated) region}
\label{fig:quietcubeskel}
\end{subfigure}
\caption[Active and quiet region \texttt{FilFinder} velocity-coded skeletons]{0th moment maps of the column density cubes with the identified filamentary structure, colour coded by velocity, overlaid on top. 2.85~pc scale bars are shown in the top right corners.}
\end{figure*}
To estimate the flow rates associated with the identified filamentary structures in each of the regions we follow the approach outlined in \cite{2024Wells} based on \cite{2020Beuther}.  The mass flows rates $\dot{M}$ are estimated as
\begin{equation}
    \dot M = \Sigma \cdot \Delta v \cdot w \cdot \frac{1}{\tan(i)}
    \label{flow_equation}
\end{equation}
where $\Sigma$ is the surface density in units of g cm$^{-2}$, taken from the data cube directly, $\Delta v$ is the velocity difference in kms$^{-1}$, calculated in different ways depending if its along, onto or polar (see Sect. \ref{sect:veldiff}). $w$ is the width of the area along which the flow rate is measured in AU, we use two pixels (0.58 pc). The final values of $\dot M$ are converted to M$_{\sun}$yr$^{-1}$. The correction factor of $\tan(i)^{-1}$ is for the unknown filament inclination, based on the discussion in \cite{2024Wells}. Here, we do not apply that correction directly but we investigate inclination separately in Sect. \ref{sect:inc}.

\subsection{Filament identification}
\label{sect:filfind}

\begin{figure}[]
    \centering
    \includegraphics[width=0.49\textwidth, height=0.3\textheight]{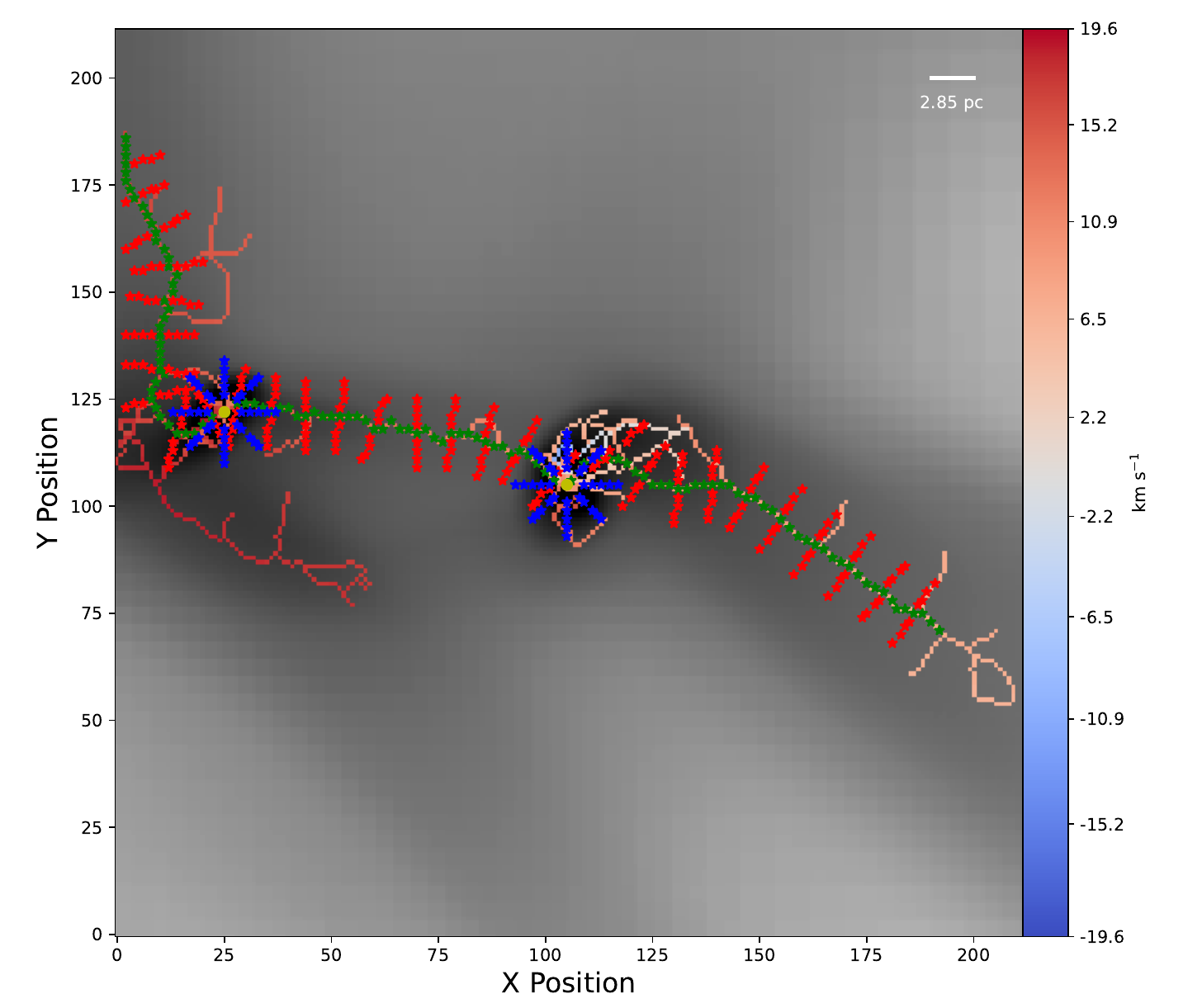}
    \caption[Along, onto and polar points demonstration]{0th moment map of the Active column density cube with the identified filamentary structure, colour coded by velocity, overlaid on top. Green, red and blue points indicating the different types of flow rate, green is along the filamentary structure, red is onto, and blue is polar around the clumps. 2.85~pc scale bar in the top right corner. }
    \label{fig:starpoints}
\end{figure}
We identify filaments in the PPV cubes using \texttt{FilFinder} \citep{2015Koch}. Specifically, we make use of \texttt{FilFinder}'s new 3D identification technique (to be presented in E. Koch et al. in preparation), which is also used for 3D filament identification in \citet{Zucker2021} and \citet{Mullens2024}. \texttt{FilFinder} in 3D uses similar morphological operations to the previous 2D version, namely using adaptive thresholding to identify locally bright structure over a large dynamic range. One key change is \texttt{FilFinder}'s use of the \texttt{skan} package to improve efficiency to handle 3D skeletons structures \citep{NunezIglesias_2018_skan}.

We use the following steps and parameters to define the filaments investigated in the subsequent analyses. First, we create a binary filament mask using a local threshold (\texttt{adapt\_thresh}) and only keep structures above a minimum surface density (\texttt{glob\_thresh}) with a minimum number of contiguous pixels (\texttt{min\_size}) to minimise spurious isolated peaks. The resulting mask is skeletonised to produce the filament spines and structure for further analysis and pruning of spurious branches on the skeleton. Table \ref{filfindparameters} shows our choice of these key parameters for the different cubes we analyse. Lastly, we note that \texttt{FilFinder} is optimised to work on elongated, filamentary structures with aspect ratios of 3:1; the masking and pruning operations described above naturally removes compact and isolated structures, though we note that isolated compact structures without surrounding filamentary structures are not found in the simulated cubes we analyse. For our analysis, we define the location of the filament and its extent using the pruned skeletons produced by \texttt{FilFinder}. 

\begin{table}[]
    \centering
    \caption[\texttt{FilFinderPPV} parameters]{\texttt{FilFinderPPV} parameters}
    \begin{tabular}{llll}\hline
       Cube  &  $adapt_{thresh}$ & $glob_{thresh}$ & pruning (px)\\\hline\hline
       Active & 13 & 0.0125 & 0 \\
       Quiet & 21 & 0.0075 & 50\\\hline    
    \end{tabular}
    \label{filfindparameters}
\end{table}

\begin{figure*}[h!]
    \centering
    \includegraphics[width=0.95\textwidth, height=0.3\textheight]{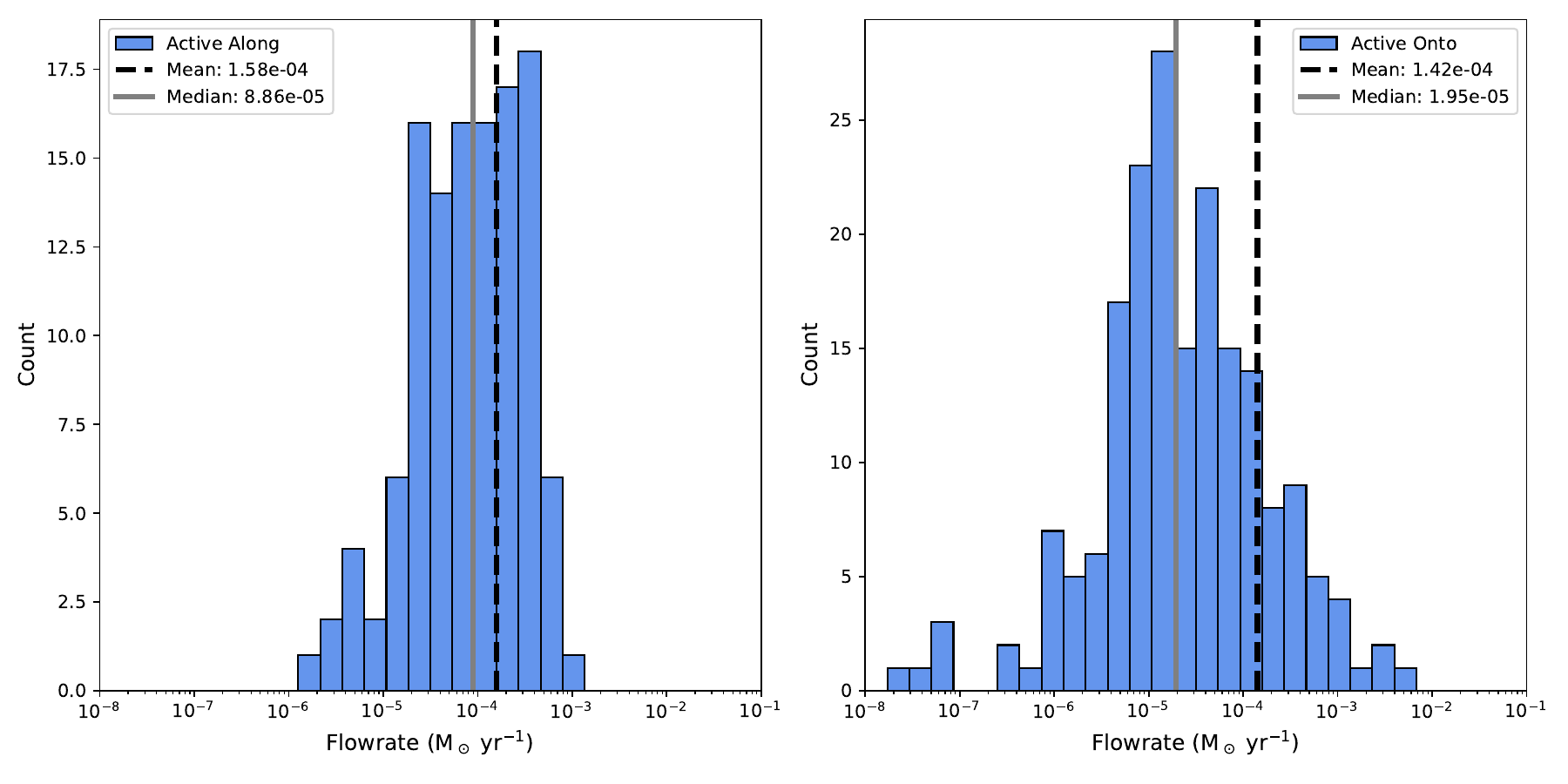}
    \caption[Active cube flow rates statistics]{Distributions of flow rates \textit{Left:} along filaments \textit{Right:} onto the filaments in the active cube.}
    \label{fig:active_stats}
\end{figure*}
\begin{figure*}[ht]
    \centering
    \includegraphics[width=0.95\textwidth, height=0.3\textheight]{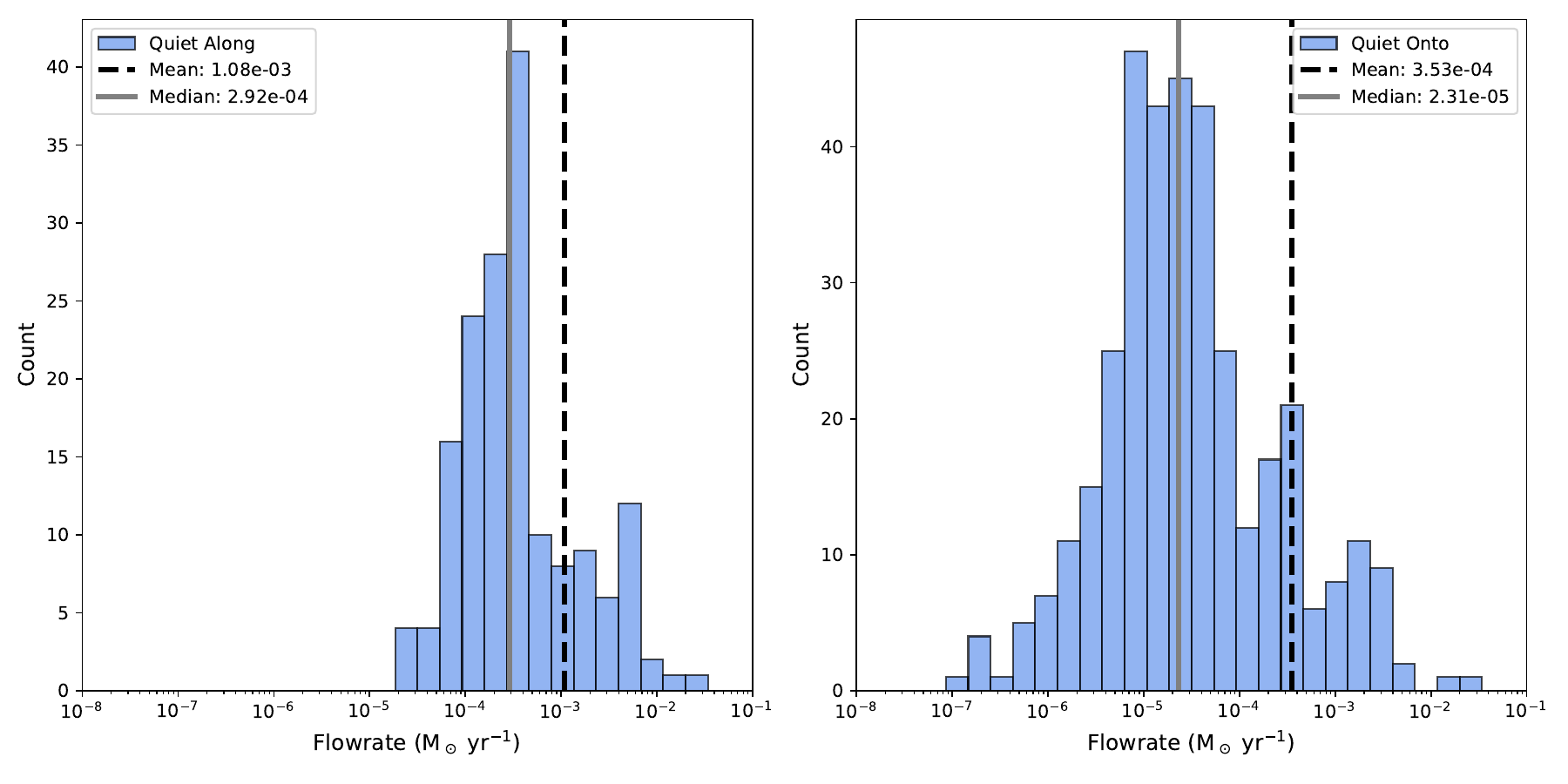}
    \caption[Quiet cube flow rate statistics]{Distributions of flow rates \textit{Left:} along filaments \textit{Right:} onto the filaments in the quiet cube.}
    \label{fig:quiet_stats}
\end{figure*}

\subsection{Velocity difference}
\label{sect:veldiff}
\begin{figure*}[ht]
    \centering
    \includegraphics[width=0.95\textwidth, height=0.5\textheight]{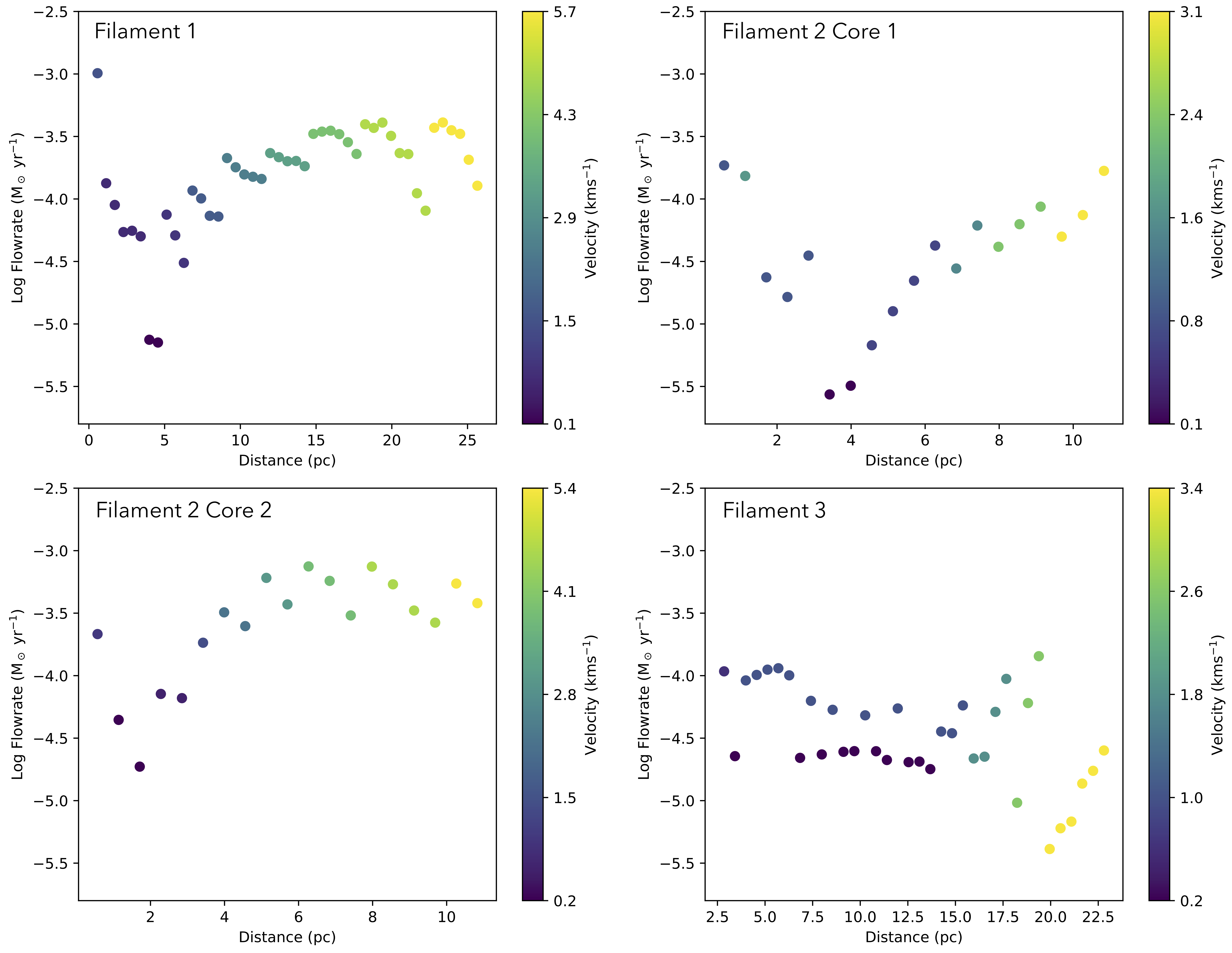}
    \caption[Flow rate vs. distance relation for the active cube]{Distance vs flow rate, in log space, relationship for each filament in the active cube (filaments are labelled in Fig. \ref{fig:activecubeskel}), colour coded by velocity difference.}
    \label{fig:active_along}
\end{figure*}
For this analysis we investigate three different types of flow fields and rates: the flow of material along filamentary structures, towards the central cluster-forming clumps; the flow of material from the environment onto the filamentary structures; and the radial inflow around the cluster (designated "polar") towards the forming clumps. Each of these flow fields needs a slightly different method for calculating the velocity difference, these are outlined in the following sections.

The flow rate equation used in this analysis works on assumptions regarding the cause of the velocity differences. 
Rotational signatures cannot be ruled out, however when looking along the filament we minimise effects from the rotation of the filament. As we move towards the central clumps, where local rotation or shear may occur, previous studies (e.g.,~\citealt{2020AXu, 2024Xu}) show that, such motions contribute minimally.

\subsubsection{Along}
Moving along the filament we calculate the velocity gradient between each point on the filament spine and the "hub" where the filaments converge. These “hubs” are typically cluster-forming regions and we will refer to them as clumps throughout this work. 
In Fig. \ref{fig:starpoints}, the green points indicate the positions along the filament at which we calculate the flow rates. We use between 30-50 points per filament, at 0.58~pc (2~px) distance increments. We use the velocity value identified by \texttt{FilFinder} along the filament and calculate the difference between that value and the clump velocity which is estimated by fitting a Gaussian to the spectrum at the central pixel of the clump.

\subsubsection{Onto}
To calculate the flow rate onto the filaments, we take four positions on either side of the filament, (see the red points in Fig. \ref{fig:starpoints}), at 0.58~pc distance increments (2 pixels). Here the velocity difference is calculated in reference to the point where the perpendicular points meet the filament. For the perpendicular points we use the Gaussian fit to the spectrum method to get the velocity at the point, and the velocity on the filament is the same as above.

\subsubsection{Polar}
Clumps are fed by a number of filaments.  We calculate the flow rates radially outwards from the core, along these converging filaments to include the contributions from each of the primary and feeder filaments (filamentary sub-structures), as well as contributions coming from the environment surrounding the clump in the absence of any filamentary structures. Here we define feeder filaments as smaller filamentary structures aiding in the flow of material either onto the primary filaments or onto the central star forming clump. 
These positions are marked by the blue points in Fig. \ref{fig:starpoints}, again at 0.58~pc distance increments. These points use the Gaussian fit to their spectrum for their velocity values and the difference is in relation to the core velocity, also calculated with this method.

\subsection{Error analysis}
\label{sect:errors}
Our flow rate equation consists of three parameters. Two are taken directly from the data cube itself, the column density and spatial resolution. The velocity values (\texttt{FilFinder} identified or Gaussian fitted) introduce the majority of the error to the final flow rate values.  The velocity resolution in the cubes is 0.8~kms$^{-1}$, so we take an estimate for the \texttt{FilFinder} skeleton identification error to be one channel, 0.8~kms$^{-1}$. As for the Gaussian fitting, we take between $\sim$~10 and 20\%, as the average error of the Gaussian fit. We conservatively assume an uncertainty of ~20\% on the estimated flow rates. We note that in real observations the uncertainties are larger because additional systematic errors from the column density estimates and projection effects come into play. These values are without error bars in our analysis because they are taken directly from the simulations.

\section{Results: Observational Method}
\label{sect:theoryresults}
\begin{figure*}[h!]
    \centering
    \includegraphics[width=0.95\textwidth, height=0.5\textheight]{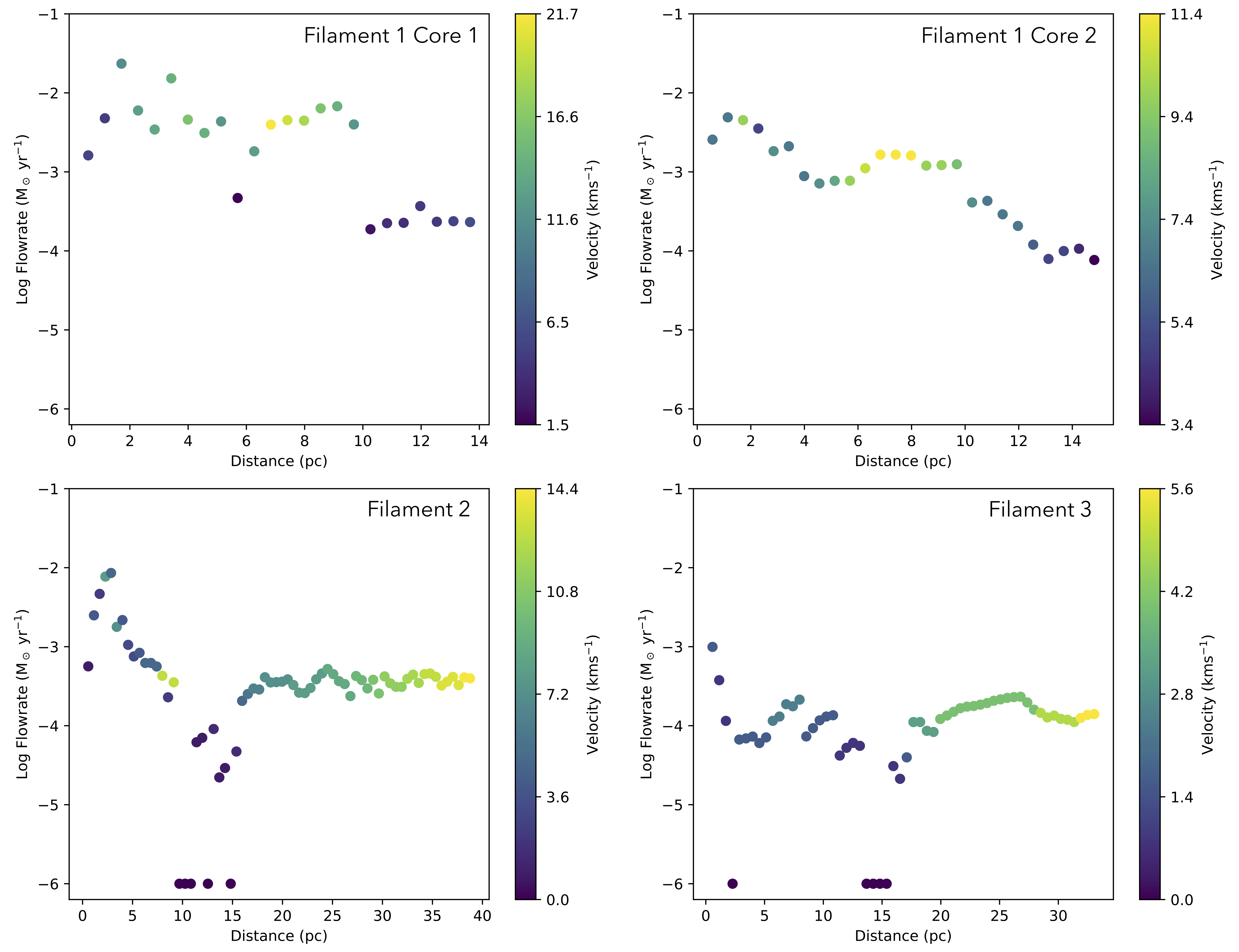}
    \caption[Flow rate vs. distance relation for the quiet cube]{Distance vs flow rate, in log space, relationship for each filament in the quiet cube (filaments are labelled in Fig. \ref{fig:quietcubeskel}), colour coded by velocity difference.}
    \label{fig:quiet_along}
\end{figure*}
In this section the flow rates calculated with the observational method presented in Sect. \ref{sect:theorymethod}.The flow rates taken directly from the 3D numerical data are presented in the following section Sect. \ref{sect:results_simulations} where we also compare them with those deduced by the observational method.  We broaden the scope of our results by comparing them with measurements from other observational programs in Sect. \ref{sect:theorydiscussion}.

Calculating the flow rates along and onto the filaments for the active and quiet regions at 0 deg inclination produce the histograms shown in  Figs. \ref{fig:active_stats} and \ref{fig:quiet_stats}. The distribution on the left panel of each figure is for flow rates along the filaments, while  the distribution on the right corresponds to flow rates onto the filaments. 

In Fig. \ref{fig:active_stats}, we present
flow rate distributions for flow rates along the filaments (left panel) and onto the filaments (right panel) in the active cube. These range between 10$^{-8}$~\flowrate and 10$^{-2}$~\flowrate with median values of 8.9$\times$10$^{-5}$~\flowrate along and 2$\times$10$^{-5}$~\flowrate onto. We also see that the active distributions have similar mean and median values within our reported errors (see Sect. \ref{sect:errors}). In Fig. \ref{fig:quiet_stats}, we see the same but for the quiet cube. Here, the range of flow rates is between 10$^{-7}$~\flowrate and 10$^{-1}$~\flowrate, with median values of 2.9$\times$10$^{-4}$~\flowrate along and 2.3$\times$10$^{-5}$~\flowrate onto.  This figure also shows that the quiet region has a significant difference between the distribution for along and onto the filaments, with their medians being separated by a whole order of magnitude. 

The distribution for the flow rates onto the filaments is wider for both the active and the quiet cubes. These values range from being right next to the filament to $\sim$ 2.5~pc away, so there is likely to be a large variation. 
We would like to note that the flow rates onto the filaments are estimated only a selected cuts across them, but that ultimately gas flows onto the filament everywhere. Therefore, the flow rates onto the filaments have to be considered as lower limits. Having only an order of magnitude difference in these individual cuts, we conclude that the flows onto are more than enough to be "feeding" the flows along the filament and towards the central clumps.

\subsection{Along}

For flows along the filaments, we analysed four different filaments in each region, focusing on flows directed towards the clumps. In the active region (Fig. \ref{fig:active_along}), the filaments exhibit a trend of increasing flow rates with distance from the clump.
This is consistent with the idea that at large scales the material is feeding whole clusters and getting closer to the central clumps this feeding can split up into several separate flows, which has also been seen on smaller scales (e.g.,~\cite{Padoan2020}).

In the quiet region (Fig. \ref{fig:quiet_along}) however, we see different trends. In two instances we see flow rates decreasing as the distance to the clump increases (see top two panels in Fig. \ref{fig:quiet_along}), and also a more constant relationship after initial peaks (potentially due to higher column density). Comparing the filament morphology in both regions reveals that these differences can be explained by the presence and distinct roles of feeder filaments alongside the main filaments analysed.
In the active region, feeder filaments primarily occur at the clump end of the main filaments. Here, the main filament splits into feeders as it approaches the hub, channelling the large flow rates across multiple paths and thereby reducing the flow rates closer to the core on the main filament. In contrast, the quiet region shows a different pattern. Feeder filaments are not concentrated near the hub but are distributed along the length of the main filaments. These feeders merge into the main filaments at various points, resulting in higher flow rates reaching the central hubs. 

This concept also accounts for the velocity peaks observed in the top panels of Fig. \ref{fig:quiet_along}, which correspond to the locations where these feeders join the main filaments. In the bottom panels of Fig. \ref{fig:quiet_along} the trends start off with high flow rates close to the clump before evening out to constant flow rates with distance. This initial peak can be attributed to the column density contributions due to the extended area around the core where the column densities are higher.

\begin{figure*}[ht!]
    \centering
    \includegraphics[width=0.78\textwidth, height=0.24\textheight]{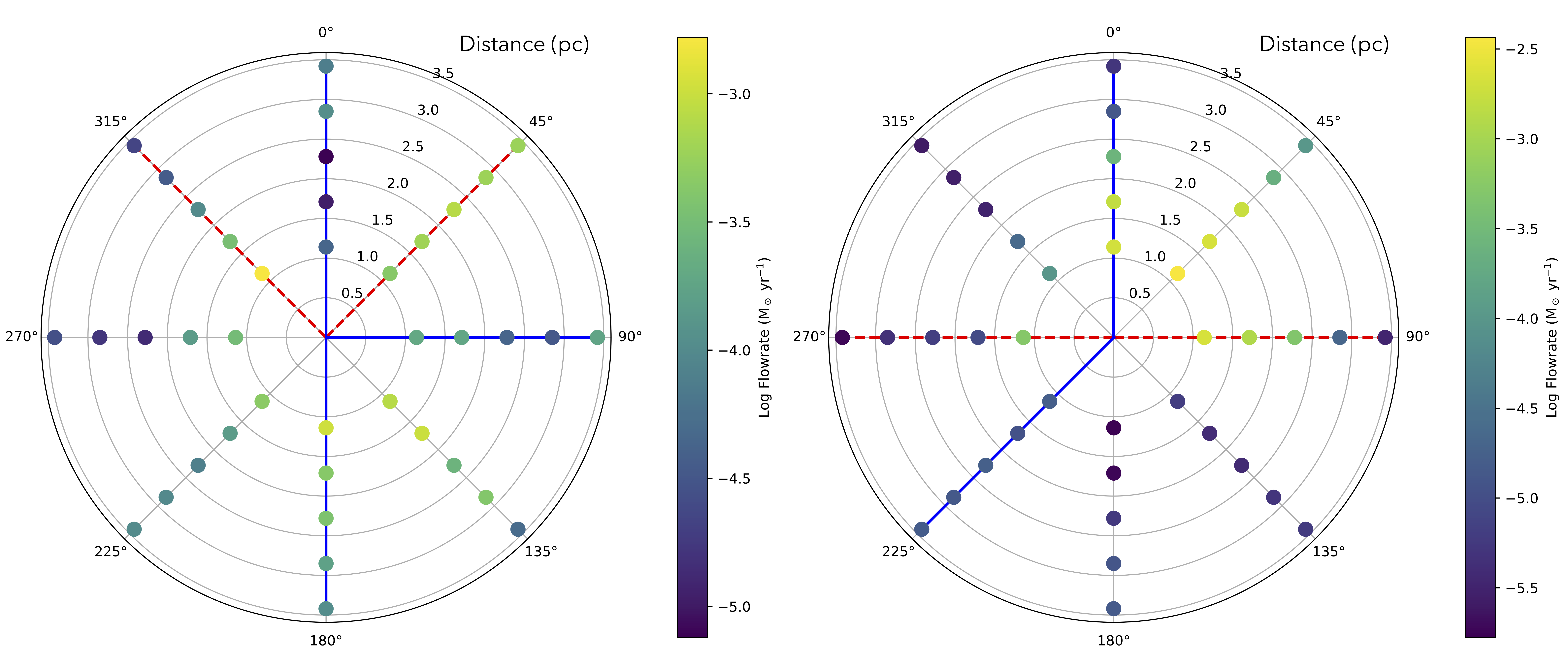}
    \caption[Active cube polar flow rate plots]{Radial distance - flow rate relationship for eight different angles around each clump in the active region. Numbers from 0.5 to 3.5 represent the distance from the centre for each of the concentric circles, in pc. Red dashed lines indicate the 'primary' filaments and blue solid lines indicate directions of the 'feeder' filaments.}
    \label{fig:active_polar}
\end{figure*}
\begin{figure*}[ht!]
    \centering
    \includegraphics[width=0.98\textwidth, height=0.20\textheight]{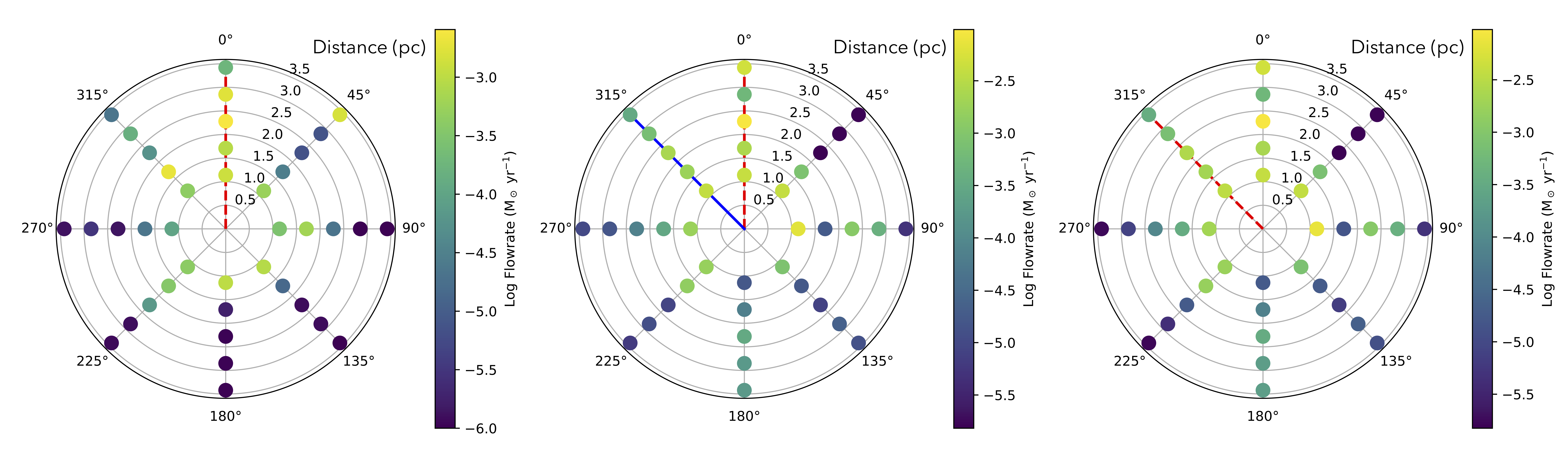}
    \caption[Quiet cube polar flow rate plots]{Radial distance - flow rate relationship for eight different angles around each clump in the quiet region. Numbers from 0.5 to 3.5 represent the distance from the centre for each of the concentric circles, in pc. Red dashed lines indicate the 'primary' filaments and blue solid lines indicate directions of the 'feeder' filaments.}
    \label{fig:quiet_polar}
\end{figure*}

\subsection{Polar}
By examining the flow rates radially around the central star-forming clumps we can identify the directions from which the largest contributions of material to the hub clump arise. Figures \ref{fig:active_polar} and \ref{fig:quiet_polar} show the flow rate values around each clump in both cubes. The red and blue lines overlaid show where the main filaments connect to the hub and where feeder filaments are.  We see that there are far more feeder filaments around clumps in the active cube, and that their contribution is significant to the flow of material onto the clump. The quiet cube, in contrast, has almost no contributions outside of its main filaments connecting to the clumps. In both Fig. \ref{fig:active_polar} and Fig. \ref{fig:quiet_polar} we see that from most angles around the clumps there is a gradient where the flow rate is increasing towards the centre which agrees with what we see in Figs. \ref{fig:active_along} and \ref{fig:quiet_along}.

Looking at many angles around each clump at small scales ($\sim$ 1 pc), in both the active and quiet regions, also shows that there is significant contribution to the flows onto the clump from the environment without the use of filamentary structures.

\subsection{Inclination}
Another important aspect of this work was the ability to investigate inclination effects on the flow rates. We look at both filamentary and galactic inclination in the following sections.
\subsubsection{Filament}
\begin{figure}
    \centering
    \includegraphics[width=0.99\linewidth]{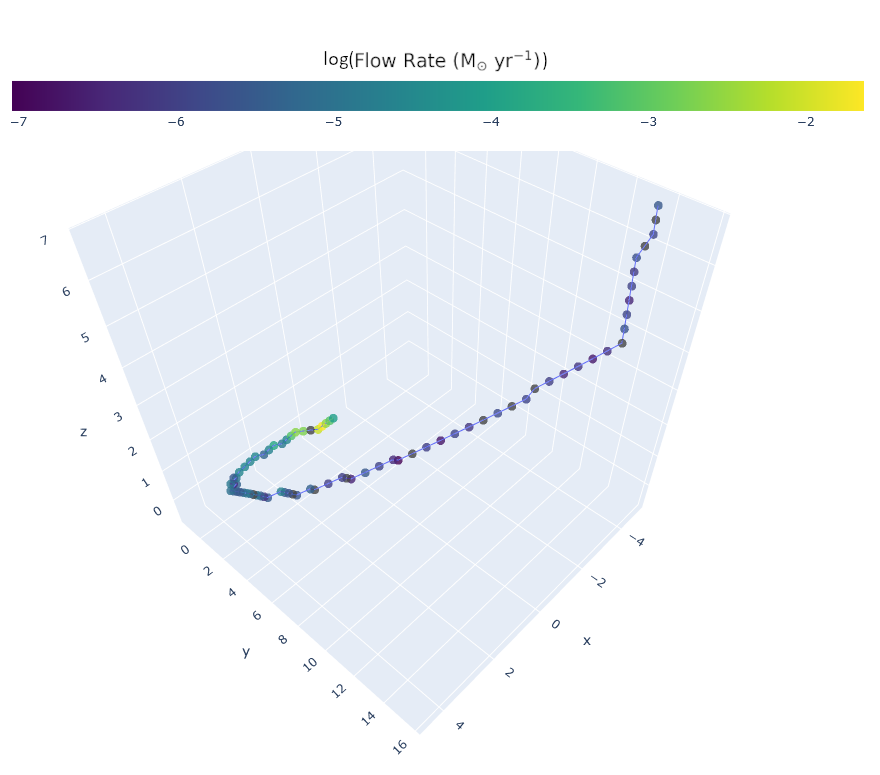}
    \includegraphics[width=0.99\linewidth]{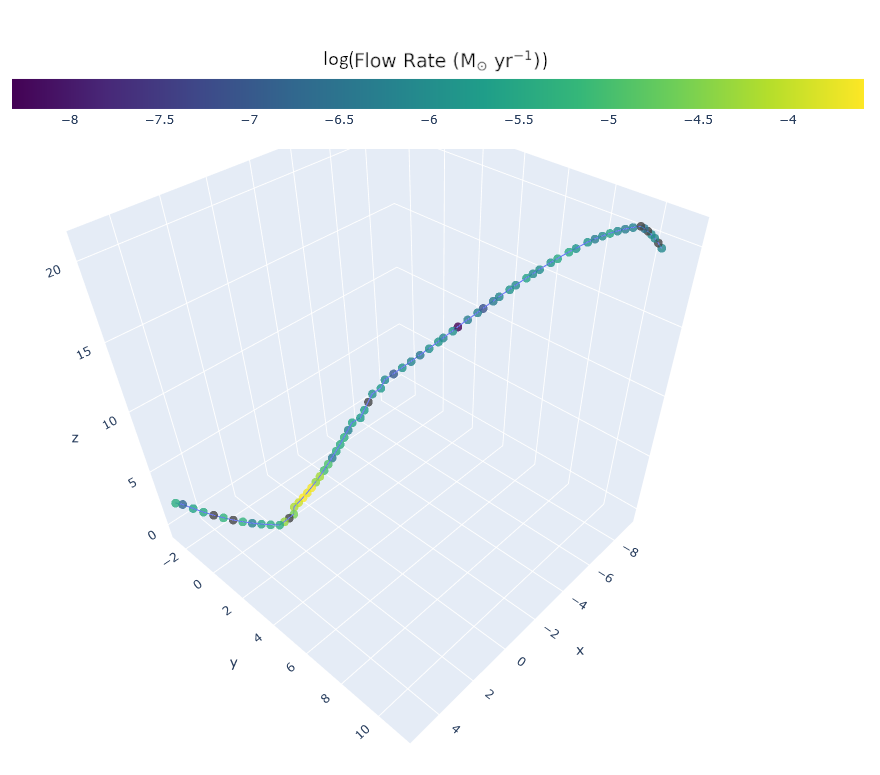}
    \caption{3D projection of the filament spines for Filaments 1 (\textit{top}) and 3 (\textit{bottom}) in the active cube. Colourbar shows the parallel flow rate at that point in the spine, grey coloured markers depict points for which the direction vector of the spine was null. The light purple plane surface shows the position of the galaxy midplane. An interactive version of this figure is available in the online version of this paper.}
    \label{fig:parallel_3D}
\end{figure}
\begin{figure}[ht!]
    \centering
    \includegraphics[width=0.5\textwidth, height=0.6\textheight]{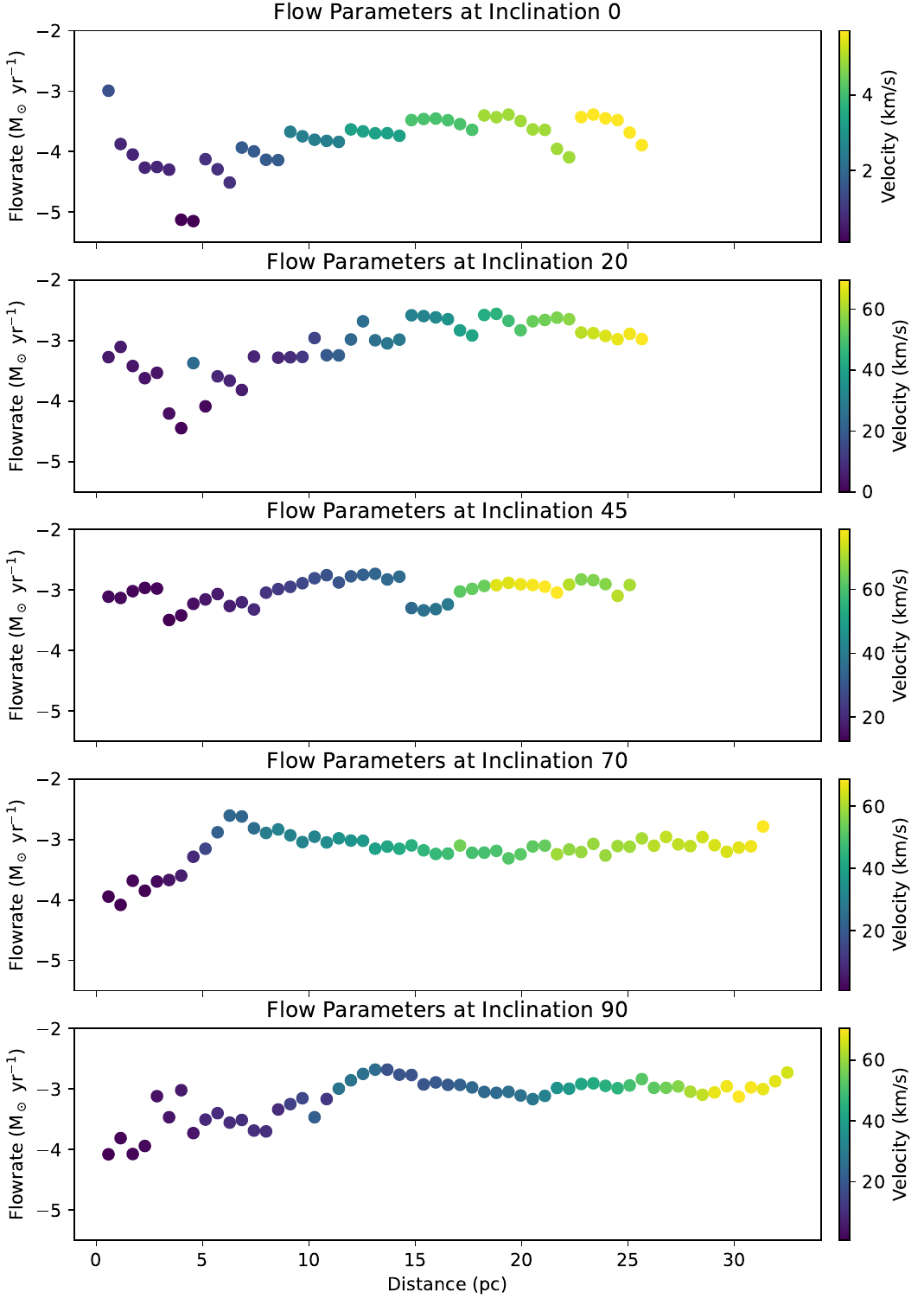}
    \caption[Inclinations distance vs. flow rate relationship]{The effect of Galactic inclination on the distance vs. flow rate relationship, in log space, for filament 1 from the active region. Results are shown for five different Galactic inclinations, 0, 20, 45, 70, and 90 degrees.}
    \label{fig:inclinations}
\end{figure}

Filament inclination with respect to the galactic plane is more complicated. We can deduce that the filament is in the plane of the sky if there is no velocity gradient, but otherwise, the inclination angle is often treated as unknown in observational studies. For example, \cite{2024Wells} estimate the effects unknown inclination values have on the final flow values. They report that with unknown inclination angle they see a larger spread in the flow rate values, with the distribution peaking close to the ``true'' flow rate. 
However, with simulation data we can now obtain an estimate for an average inclination along the filament with respect to the galactic plane. 
We measure the angles of the filaments with respect to the galactic plane by sampling direction vectors along the 3D spines of the filaments. The angle between the two vectors is easily determined via their dot product. 
We measure the circular mean both unweighted and weighted by the parallel flow rate to calculate the average inclination for the entire filament. We calculate the weighted mean using the parallel flow rate, at each point along the spine as the weight. As such, the regions closer to the forming cluster with higher flow rates are weighted more heavily by the value of the inclination.

The results presented in Sect. \ref{sect:inc} show the effects of galactic inclination where we incrementally increase the galactic inclination from 0 degrees (face on) to 90 degrees (edge on).
We see that the range of velocity differences increases in the inclined cubes relative to the face-on cube. As a result, we see trends at all inclinations shown in Fig. \ref{fig:inclinations} with typical flow rates of 10$^{-3}$~\flowrate to 10$^{-4}$~\flowrate. Both the inclination and the presence of feeder filaments work together in this region to create different effects. The unknown inclination here will have similar effects on the feeder filaments as the primary filament. On the other hand, galactic inclination rotates the filament, with the potential to reveal other feeder filaments or more details about the structures that are not apparent from other angles. 

In Figure \ref{fig:parallel_3D}, we present 3D plots of the filaments we have identified. Importantly, these show the position relative to the mid-plane of the galactic disk (shown by the light purple plane in the plot). This cluster-forming area sits 313 pc above the mid-plane of the galaxy, just within one scale height of the galaxy. 
For Filament 1, we measure an unweighted average of $|49.8|^\circ$ and a flow rate-weighted average of $|48.4|^\circ$. The two averages are consistent within error implying that the angle is similar along the entire filament. For Filament 3, we measure an unweighted average angle of $|10.9|^\circ$ and a flow rate-weighted average of $|0.8|^\circ$. From this, we deduce that the angle with respect to the plane is not consistent along the entire filament, with the highest flow rate areas closest to the clump tending to be more parallel to the plane. However, we note that both of these filaments are not coplanar with the galaxy midplane as they sit more than 300 pc above the mid-plane, as shown in Figure \ref{fig:parallel_3D}.

Taking into account the flow rate weighted average inclination angles we estimated from the simulation, we can compare the affect of the $1/\mathrm{tan}(i)$ inclination correction factor for on the the distribution of the flow rates for these two filaments. For Filament 1, we find $1/\mathrm{tan}(i)\sim$~0.8, which means that our observed flow rates for this filament are slightly overestimated when inclination angle is unknown. 
For Filament 3, however, the inclination factor is $1/\mathrm{tan}(i)\sim$~72, meaning that our estimates are underestimated by over an order of magnitude. This is not surprising when looking at Fig. \ref{fig:active_along} we see the bottom right panel, Filament 3, has the lowest flow rates of the region. In general, it’s important to consider unknown inclination angles when measuring and discussing observational flow rates, as they are a key factor in accurately interpreting the results and understanding the flow behaviour.

\begin{figure}
    \centering
    \includegraphics[width=1.0\linewidth]{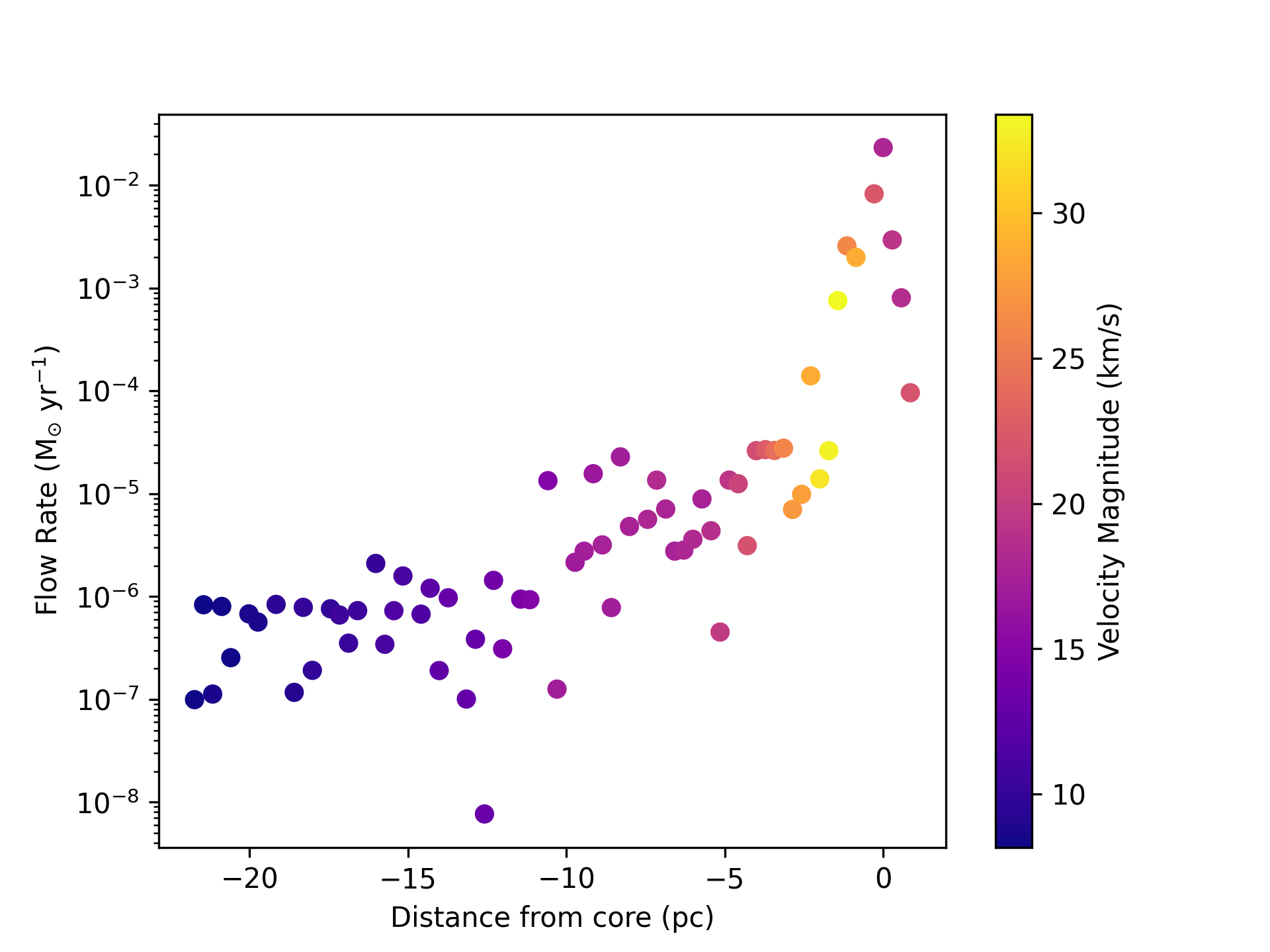}
    \includegraphics[width=1.0\linewidth]{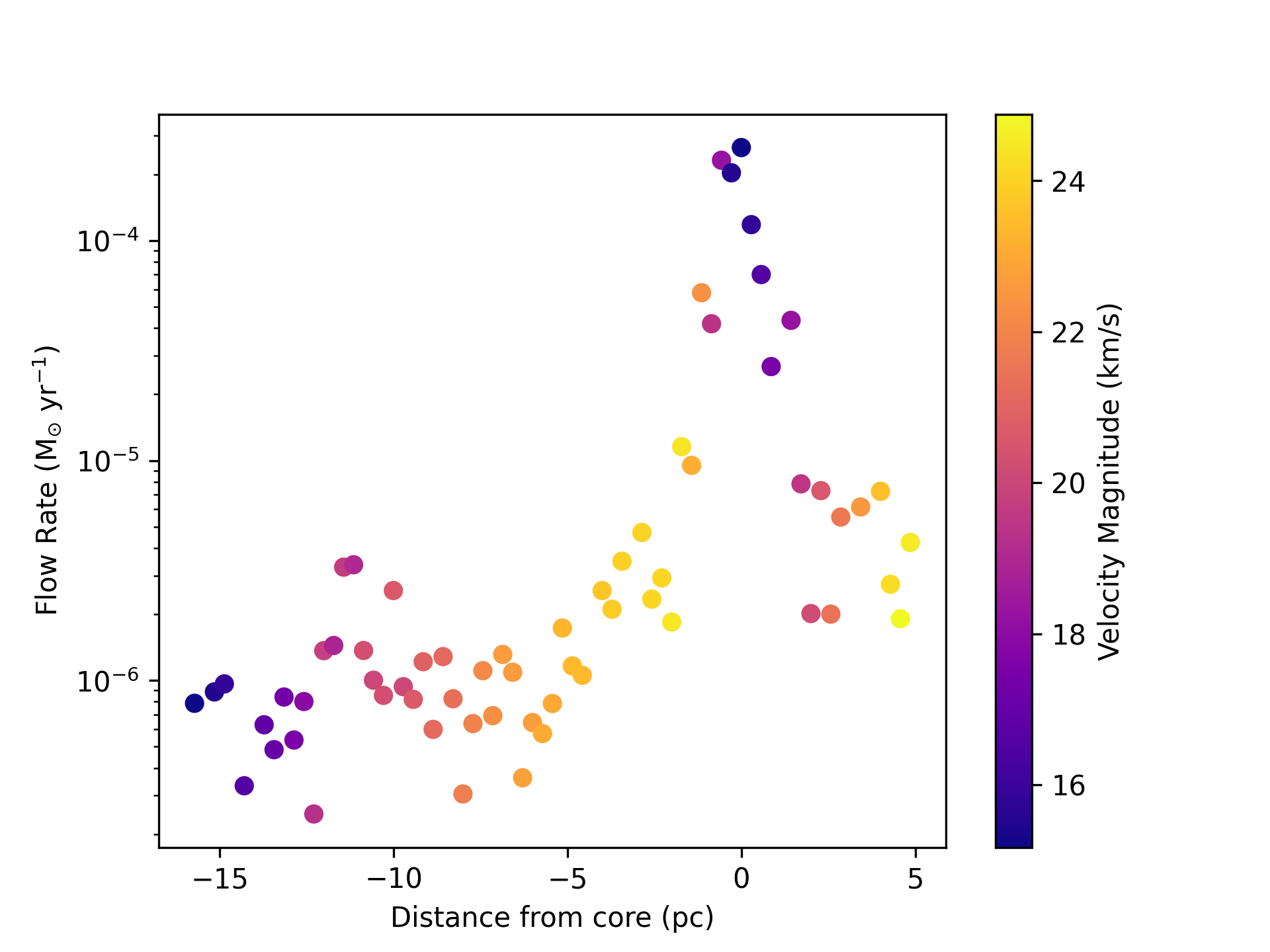}
    \caption{Flow rates along the filaments in the active cube, with colourbar representing the velocity magnitude at that point. \textit{Top:} 3D filament corresponding to filament 1 in Figure \ref{fig:activecubeskel}, feeding the central, massive clump. \textit{Bottom:} the 3D filament representing Filament 3 in Figure \ref{fig:activecubeskel}, feeding the non-central clump. Negative values represent positions leftwards of the clump, i.e. lower values of x and y. On the x-axis 0 corresponds to the clump position.}
    \label{fig:parallel_scatter}
\end{figure}
\subsubsection{Galactic}
\label{sect:inc}

The inclination angle ---both of the filament and galactic disc--- directly influences the magnitude of the velocities measured in observations. Although the galactic disc inclination angle can typically be measured in external galaxies, inclination effects of filaments in the Milky Way are often poorly constrained. This uncertainty motivates us to  take an in-depth look at the effects of Galactic and filament inclination on our estimated measurements of the flow rates along filamentary structures towards the central clump. 

To pursue this goal, we take one filament and clump structure from the active region to focus on throughout the inclinations of 0 (face on), 20, 45, 70 and 90 (edge on) degrees. The results are shown in Fig \ref{fig:inclinations}. We see overall throughout the five inclinations the relationship between distance and flow rate varies by a factor 10. Looking at 0 and 20 degrees we see initial peaks and then a small drop before steadily increasing by an order of magnitude, for 0 degrees to $\sim$~10$^{-3.5}$~\flowrate and for 20 degrees to $\sim$~10$^{-2.5}$~\flowrate. At 45 degrees the values are almost constant around 10$^{-3}$~\flowrate. At the final two inclinations, 70 and 90 degrees, we see the values start lower, at 10$^{-4}$~\flowrate, than the previous inclinations before increasing, peaking around 10$^{-2.5}$~\flowrate and plateauing to become constant at larger distances from the clump, around 10$^{-3}$~\flowrate. Flow rates onto the filament were also calculated at each inclination, the median values for each inclination are all on the order of 10$^{-5}$~\flowrate, ranging between 1-7$\times$10$^{-5}$~\flowrate with a small increasing trend from 0 through to 90 degree inclination.

\section{3D Simulation Flow Rates - and Comparison with the Observational Method}
\label{sect:results_simulations}
\begin{figure*}
    \centering
    \includegraphics[width=0.49\linewidth]{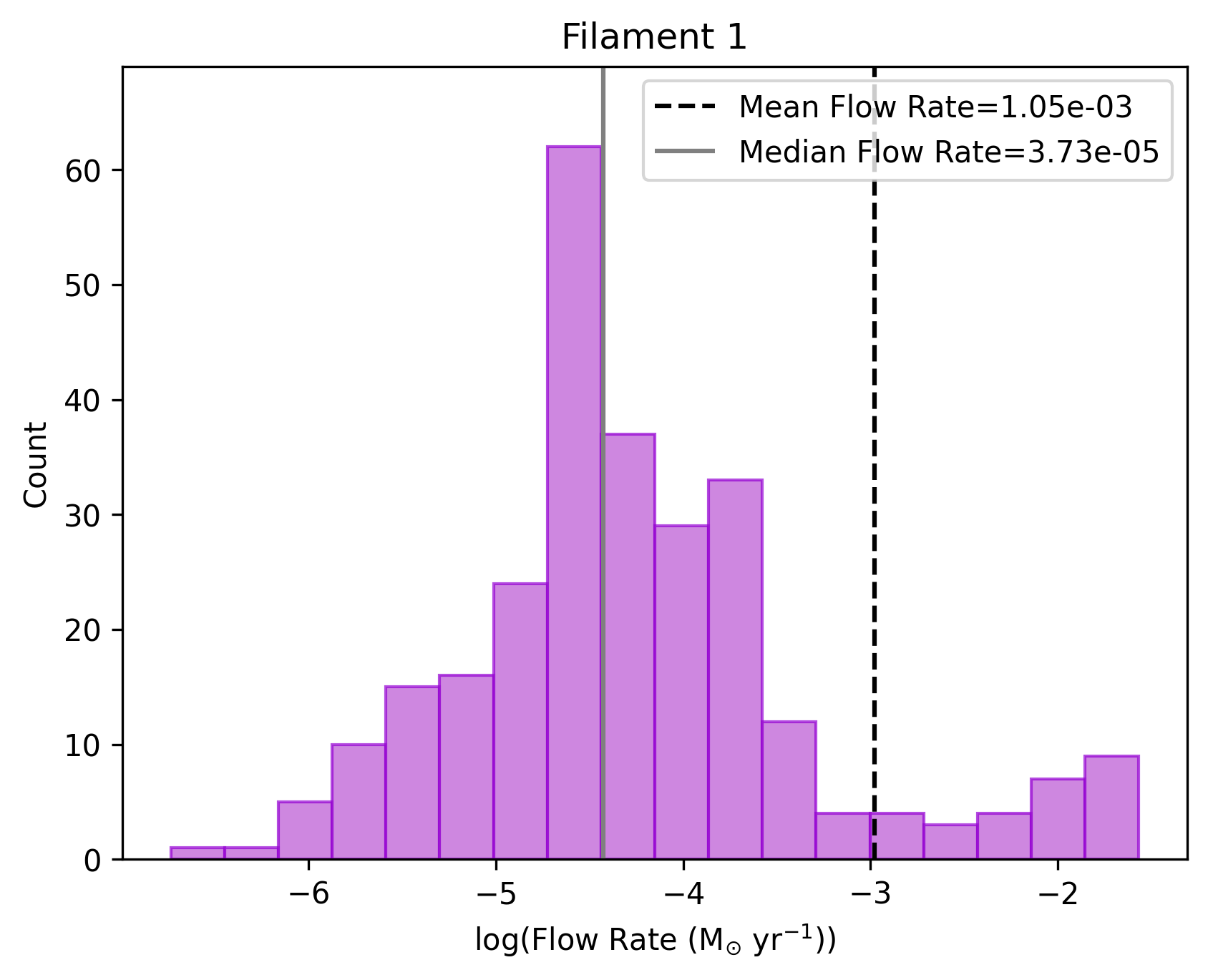}
    \includegraphics[width=0.49\linewidth]{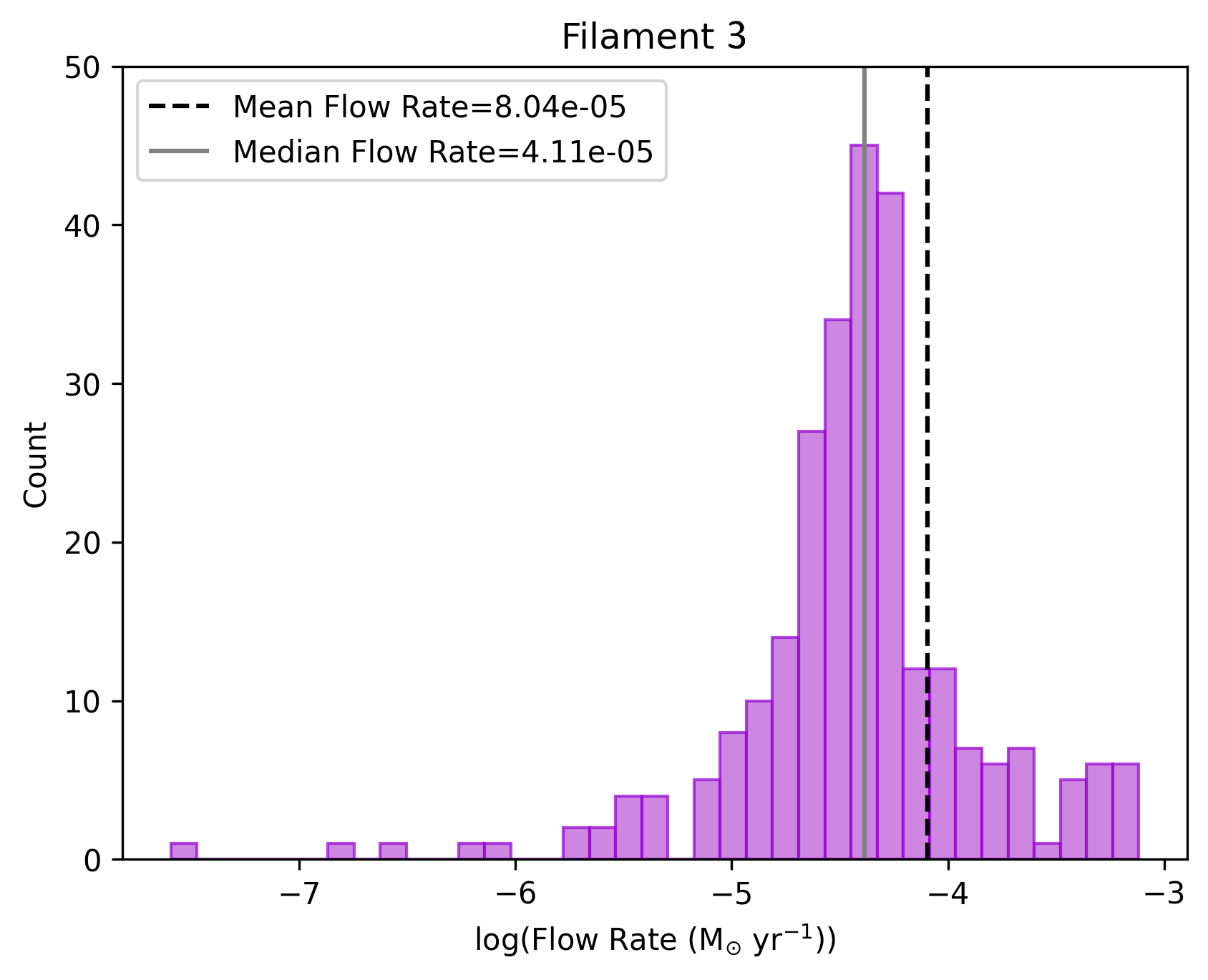}
    \caption{Distributions of all the perpendicular flow rates measured at 0, $\pi/2$, $\pi$, and $3\pi/2$  for the 3D filament spines in the active region. Black dashed and grey solid lines sit at the mean and median flow rate values, respectively, with numbers shown in the legend. \textit{Left:} Filament 1, which feeds the central, massive clump. \textit{Right:} Filament 3, which feeds the non-central clump.}
    \label{fig:perpflows}
\end{figure*}

A particularly interesting aspect of these observed measured flow rates for each filament is that we can directly compare them to their ``true'' values from the full, 3D simulation data. In this section we provide these ``true'' values by taking an approximation of the filament in 3D.  This is achieved by masking the (x,y) values contained in the skeleton and finding the peak density along the z-axis at each point. Within the region explored, the maximum density in z will represent the third dimension of the spine of the filament, assuming the spine is aligned with the dense ridge of the filament (as is done in filament profile fitters, such as RadFil in \citet{ChenZucker2018}). We visually check for connectivity of this filament, ensuring that the z-values contribute to a continuous filament in 3D projections of the gas density. 

With an extracted 3D approximation of the filament in hand, we project cartesian velocity fields onto the axis of the filament to measure the parallel components. The perpendicular vector is then the vector subtraction of the original velocity vector and the filament's parallel axis, and provides us with perpendicular components of the velocity field. The cross product of the two existing vectors contributes to the second perpendicular vector, allowing us to measure flows along four directions onto the filament. Flow rates onto the filament, perpendicular to its spine, are computed in 4 directions (0, $\pi/2$, $\pi$, $3\pi/2$). Each measurement is taken as the average flow rate from a vector extending 2.8 pc away from the spine of the filament.
The flow rate of gas moving onto a single fluid element can be expressed with the density, velocity and the area being measured. For a single fluid cell, this is
\begin{equation}
    \dot{m} = \rho \vec{v}_nA
\end{equation}
\noindent where $\rho$ is the volume density of gas moving through area A at a velocity normal to the surface $\vec{v}_n$. Measuring a flow rate, as opposed to tracking the change in mass over multiple timestamps, allows us to separate between parallel and perpendicular flow rates (i.e. along and onto the filament, respectively) while being able to neglect any changes that may be due to the changing morphology of the filamentary structure itself due to the dynamics in the larger galactic environment. With this approach, we measure the following flow rates on the two main filaments identified in the active cube that feed each clump. 

Figure \ref{fig:parallel_scatter} shows the flow rates measured along the spines of the filaments in the active region, corrected for the distance from the main core in the structure. The scatter points are coloured by velocity magnitude, similar to Figure \ref{fig:active_along}, using the 3D vector. 
The top plot in Figure \ref{fig:parallel_scatter} represents Filament 1, which feeds the more massive clump. Flow rates along this filament average 3$\times10^{-4}~\mathrm{M}_{\odot}~\mathrm{yr}^{-1}$, but spreads to both higher and lower flow rates than the observational methods show. While the average flow rate agrees with the mean parallel flow rate in the active cube, the larger spread in the distribution of flow rates might suggest that lower values of flow rates are overestimated.
The bottom plot in Figure \ref{fig:parallel_scatter} represents Filament 3, which feeds the smaller, less compact clump in the data. Flow rates along the filament show little spread, only 2 orders of magnitude, and average to $1.5\times10^{-5}$\flowrate. These averages agree with the median flow rate measured in our observational methods presented above. 

Figure \ref{fig:perpflows} shows the distributions of the perpendicular flow rates for both of our 3D filaments in the active region. The left panel gives a median value of 
3.7$\times10^{-5}$~\flowrate for Filament 1, the right panel gives a median value of 4.1$\times10^{-5}$~\flowrate for the Filament 3. Both values agree on order of magnitude with the observationally calculated distribution. We similarly conclude here that these values are enough to sustain the flow rates along each of these filaments.

Our simulation results for flow rate-distance relationships show an opposite relation to those measured in \cite{Padoan2020}. The flow rates from this study are shown to be increasing towards the core, while \cite{Padoan2020} find that their flow rates decrease towards the core (their Fig. 17). 
We expect that this difference is primarily due to the different scales on which the flow rates are measured, as we explore the trends across $\sim$~20~pc scales while \cite{Padoan2020} focus on the innermost 1~pc.
For this part of the work we have focused on the filaments of the active, feedback-dominated region on scales of $\sim$20 pc. Our results may imply different flow behaviour in these larger scale regions than the small-scale turbulent box simulations from \cite{Padoan2020}. At $\sim$20 pc from the central clump, the gravitational force from it is unlikely to be the dominant effect on the velocity field, whereas the innermost 1 pc is situated within the gravitational potential of the forming cluster and the region's fields will naturally be affected by the dense clump's gravitational influence.

\section{Discussion}
\label{sect:theorydiscussion}
These results provide us with several key insights on the effect that the environment plays in the morphology and kinematics of filaments and in particular upon the flow rate trends we measure.  

The largest flow rates, on the order of $\sim$~10$^{-1.5}$~\flowrate, are found along the filaments in the quiet (less feedback dominated) region (see Fig. \ref{fig:quiet_stats}). These are over an order of magnitude ($\sim$~10$^{-1.5}$~\flowrate vs. $\sim$~10$^{-3}$~\flowrate) greater than the filamentary flow in filaments formed by the collision of super bubbles - characteristic of the active, feedback-dominated, region. This suggests that the spiral arms are more effective in funnelling material than the active region.  

This is supported upon examination of Fig. \ref{fig:theorycubes} which plots the column densities of filaments in the two regions.   The column density values are, on average, higher in the spiral arm (quiescent) region, suggesting higher flow rates. Also, these higher column density regions are more spatially extended from the clumps in comparison to the active region where this intensity is concentrated on the clumps themselves. 
Both of these differences suggest that the dynamics of flows in spiral arms may play a more important role than stellar feedback in driving filament-aligned flow rates that feed gas into clumps and cluster-forming regions. Clearly many more such regions, such as the Central Molecular Zone (CMZ) or inter-arm regions, need to be examined before we can make firm conclusions and differentiate flows and structures in different galactic environments - see \citet{Pillsworth2025}

Several trends emerge from the measurements of the flow rates within both the active and quiet cubes. The active cube exhibits a clear pattern where flow rates increase with distance from the core, indicating that the flow dynamics are heavily influenced by the presence of feeder filaments near the hubs. These feeder filaments distribute the flow into multiple pathways, reducing the flow rate in the central regions of the main filaments. 
Conversely, in the quiet cube, flow rates are highest near the core and decrease with distance, suggesting a more centralised accumulation of material. The distribution of feeder filaments along the main filaments in the quiet cube contributes to this pattern, as these feeders progressively merge, allowing more material to reach the clump without significant distribution away from the central core.

The order of magnitude difference in median flow rates also suggests a more dynamic process is taking place closer to the central clumps in the active cube, driven by the presence of multiple feeder filaments. In contrast, the higher median flow rate in the quiet cube indicates that even in regions of less feedback, a significant amount of material is still funnelled towards central clumps.

\subsection{Comparison with observational programs}
Whilst we have seen that using our observational method on the simulated data gives results that agree with the values derived directly from the models themselves, it is also important to see how the results compare with previous observational studies. \cite{2010Schneider} described the impact of filamentary structures on star formation with an in-depth study of the DR21 region, and show that flows onto the primary filaments can be enough to sustain them and their flow of material. 

For comparison to observational studies on similar scales to these simulations, we turn to \cite{2020Beuther} and \cite{2024Rawat}. Starting with \cite{2020Beuther}, they look at the flow rates in the cloud surrounding IRDC G28.3 at distances up to $\sim$~15~pc, similar to some of the filaments we look at here. Their results are on the order of 10$^{-5}$~\flowrate, appear constant over the extent of the cloud. \cite{2024Rawat} identify and analyse six filaments in the GMC G148.24+00.41. The six filaments have lengths between 14-38~pc and they calculate flows rates between 10$^{-4}$~\flowrate and 10$^{-5}$~\flowrate.

There are many other works including flow rate analysis along filaments or filamentary structures on different scales ( e.g.,~\citealt{2013Kirk, 2014Henshaw, 2024Wells}) that give results in line with what we see both in our observationally calculated results and simulation-based results.

\cite{2020Kumar} emphasised the role of hub-filament systems (HFSs) in star formation, concluding that hubs can trigger and drive longitudinal flows along the filaments in their systems, this fits in nicely with the idea of feeder filaments and the two roles we have detailed in this work.

\cite{2024Zhang} look at filamentary sub-structure on much smaller scales than ours; $\sim$ 0.17~pc long.  They report flow rates, at the higher end of the range we see here, between $\sim$~1.8~$\times$~10$^{-4}$~\flowrate and $\sim$~1.2~$\times$~10$^{-3}$~\flowrate. The different scales of their filaments is a key factor here for comparison with this work, which are around 20~pc in length.  The authors also measure the flow rates within areas corresponding to the smallest scales covered by our simulations. Their values appear at the upper end of our distributions, and additional small scale effects, e.g., additional gravitational attraction, may increase their values. 

These comparisons suggest that while broad trends and values are consistent across studies, the detailed morphology and arrangement of filaments are key factors in the flow dynamics of star-forming regions, and it is of the upmost importance that we understand them. Future studies covering both larger and smaller spatial scales are needed to explore how these parameters vary with simulations and observations to shed further light on these issues. 

\section{Conclusions}
\label{sect:theoryconc}
Our analysis of the flow rates in different environmental conditions from a simulated Milky Way-type galaxy by \cite{ZhaoPudritz2024} provides significant insights into the dynamics of filamentary structures in different star-forming environments. The use of \texttt{FilFinder} identification techniques allowed us to extract and analyse filaments, revealing differences in flow patterns between these regions. Overall we see flow rates on the order of 10$^{-4}$\flowrate and 10$^{-5}$\flowrate which are in good agreement with observations (e.g.,~\citealt{2013Kirk,2014Henshaw,2020Beuther,2024Zhang,2024Wells}).  
The key take away points from this work are as follows:
\begin{itemize}
    \item Values for the observationally calculated flow rates along individual cuts onto the filaments are lower than those along the filament, the cumulative flow rates summed from these individual cuts along the filaments is enough to sustain the flow rates we see along the filaments.
    \item In the active, feedback-dominated, region, observational flow rates tend to increase with distance from the core, a pattern explained by the presence of multiple feeder filaments, within 1-2 pc of the central clump, distributing the flow into various paths onto the central clump.
    \item The quiet, more spiral arm like, region displays higher observational flow rates towards the core, suggesting a more centralised accumulation of material. The progressive merging of feeder filaments into main filaments in these regions supports a sustained material flow towards the central clumps.
    \item Radially around the clumps we identify the primary filaments along with the presence or absence of feeder filaments depending on the region. The primary filaments align with the directions of the largest radial contributions to the flow of material onto the clump.
    \item Feeder filaments play distinct roles depending on the environment. In regions with higher feedback, they channel material from the primary filaments to feed the clumps. In contrast, in environments with less feedback, feeder filaments directly supply material from the surroundings to the primary filaments themselves.
    \item Taking the average filament inclination for two of the filaments in the active region we discuss how the observationally calculated values are slightly overestimated for filament 1, the filament feeding the more massive clump, and they are underestimated by a factor of $\sim$72 for filament 3, feeding the smaller clump in the region.
    \item Our method for observationally estimating the flow rates produce results in agreement with those directly from the simulation, with similar statistics and distributions, giving us confidence in the values we could obtain using this method on observational data whilst noting the effects of inclination and projection.
    \end{itemize}
Finally, our findings align numerically with observational studies, highlighting the critical role of filamentary structures in star formation. The differences in flow dynamics and filamentary structures underscore the importance of feeder filaments in shaping the star-forming environment. Future work should focus on refining simulations to match more observational scales and exploring the impact of inclination on observed flow rates.
\begin{acknowledgements}
EWK acknowledges support from the Smithsonian Institution as a Submillimeter Array (SMA) Fellow. RP acknowledges support from an NSERC CGS-D scholarship. Some of the computational resources for this project were enabled by a grant to REP from Compute Canada/Digital Alliance Canada and carried out on the Cedar computing cluster.  REP is supported with a Discovery grant from NSERC Canada. 

\end{acknowledgements}
\bibliographystyle{aa} 
\bibliography{wells_2024.bib}
\end{document}